\documentclass[conference]{IEEEtran}
\ifCLASSINFOpdf
\else
\fi
\usepackage{multirow}
\usepackage{pifont} 
\usepackage{graphicx}
\usepackage{booktabs}
\usepackage{hyperref}
\usepackage{url}
\usepackage{doi}
\usepackage{listings} 
\usepackage{xcolor}   
\usepackage{tabularx} 
\usepackage{ragged2e} 

\makeatletter
\newcommand{\linebreakand}{%
  \end{@IEEEauthorhalign}
  \hfill\mbox{}\par
  \mbox{}\hfill\begin{@IEEEauthorhalign}
}
\makeatother

\begin{document}
%
\title{When Security Meets Usability: An Empirical Investigation of Post-Quantum Cryptography APIs}

\author{
    \IEEEauthorblockN{Marthin Toruan}
    \IEEEauthorblockA{Royal Melbourne Institute of Technology\\
        s4075803@student.rmit.edu.au}
    \and
    \IEEEauthorblockN{R.D.N. Shakya}
    \IEEEauthorblockA{University of Moratuwa\\
    shakyardn.26@uom.lk   }
    \and
    \IEEEauthorblockN{Samuel Tseitkin}
    \IEEEauthorblockA{ExeQuantum\\
        sam@exequantum.com}
    \linebreakand 
    \IEEEauthorblockN{Raymond K. Zhao}
    \IEEEauthorblockA{ExeQuantum\\
        raymond@exequantum.com}
    \and
    \IEEEauthorblockN{Nalin Arachchilage}
    \IEEEauthorblockA{Royal Melbourne Institute of Technology\\
        nalin.arachchilage@rmit.edu.au}
}

	

%


\IEEEoverridecommandlockouts
\makeatletter\def\@IEEEpubidpullup{6.5\baselineskip}\makeatother
\IEEEpubid{\parbox{\columnwidth}{
		Symposium on Usable Security and Privacy (USEC) 2026 \\
		27 February 2026, San Diego, CA, USA \\
		ISBN 978-1-970672-07-7 \\
		https://dx.doi.org/10.14722/usec.2026.23076 \\
		www.ndss-symposium.org, https://www.usablesecurity.net/USEC/
}
\hspace{\columnsep}\makebox[\columnwidth]{}}

\maketitle

\begin{abstract}
Advances in quantum computing increasingly threaten the security and privacy of data protected by current cryptosystems, particularly those relying on public-key cryptography. In response, the international cybersecurity community has prioritized the implementation of Post-Quantum Cryptography (PQC), a new cryptographic standard designed to resist quantum attacks while operating on classical computers. The National Institute of Standards and Technology (NIST) has already standardized several PQC algorithms and plans to deprecate classical asymmetric schemes, such as RSA and ECDSA, by 2035. Despite this urgency, PQC adoption remains slow, often due to limited developer expertise. Application Programming Interfaces (APIs) are intended to bridge this gap, yet prior research on classical security APIs demonstrates that poor usability of cryptographic APIs can lead developers to introduce vulnerabilities during implementation of the applications, a risk amplified by the novelty and complexity of PQC. To date, the usability of PQC APIs has not been systematically studied. This research presents an empirical evaluation of the usability of the PQC APIs, observing how developers interact with APIs and documentation during software development tasks. The study identifies cognitive factors that influence the developer's performance when working with PQC primitives with minimal onboarding. The findings highlight opportunities across the PQC ecosystem to improve developer-facing guidance, terminology alignment, and workflow examples to better support non-specialists.
\end{abstract}



\section{Introduction}
\label{sec:introduction}
Quantum computing (QC) presents a paradigm shift in computational power, most notably in its ability to solve cryptographic problems that are computationally infeasible for classical machines \cite{gill2022quantum}. Central to this advantage is Shor's algorithm \cite{Shor1994, Shor1997}, which poses a direct threat to widely used public-key encryption standards like Rivest-Shamir-Adleman (RSA) \cite{Soni2018RSAClassicalQuantum, Handa2024}. To understand the magnitude of this threat, one can compare a classical supercomputer to a burglar who attempts to crack a lock by checking one number at a time--a process that could take billions of years for 2048-bit encryption. In contrast, a quantum computer takes advantage of the principles of quantum mechanics to evaluate vast combinations simultaneously, theoretically picking the same lock in a matter of hours \cite{Gidney2021RSAfactorquantum}.

This danger has moved beyond abstract theory, as active research continues to reduce the resources required to mount such attacks. For example, recent optimizations of windowed arithmetic circuits have reduced the total gate count for factoring by approximately 3.4\% \cite{Luongo2025quantumrsa}. Although this may seem small, it certainly lowers the barrier to practical quantum cryptanalysis, indicating that widely deployed algorithms will eventually become vulnerable \cite{Bernstein2009PQC}.

The implications of this vulnerability extend far beyond digital data theft; they introduce serious physical risks. If encryption is broken, attackers could bypass security protocols in Operational Technology (OT) and National Critical Infrastructure. In a practical scenario, this capability would allow adversaries to manipulate valve controls in water treatment facilities, trigger shutdowns in national energy grids, disrupt railway transportation networks, or sabotage manufacturing production lines \cite{cisa2022quantum}.

Compounding this physical risk is the immediate strategic threat known as ``Harvest Now, Decrypt Later'' (HNDL) \cite{Singh2024HNDL, Ali2021AnalysisPQC}. Even before a fully capable quantum computer exists, attackers are already intercepting and storing large volumes of encrypted traffic from critical infrastructure systems such as water, energy, healthcare, transportation, and manufacturing, with the explicit intention of decrypting it retrospectively once quantum capabilities mature. For example, an adversary monitoring encrypted telemetry from a national power grid can archive operational logs and configuration updates that are currently protected by classical public-key cryptography. When a cryptographically capable quantum computer emerges, the attacker could decrypt the stored data, reconstruct the state of the system, and uncover structural weaknesses. Thus, the impact of HNDL attacks materializes not during interception but when future decryption becomes technically possible.

In response to these emerging threats, the National Institute of Standards and Technology (NIST) initiated its Post-Quantum Cryptography (PQC) effort in 2016, calling for the development of quantum-resistant classical algorithms. By 2025, this process had produced five standardized schemes \cite{nistSelectedAlgorithms}. In addition, NIST's PQC transition plan \cite{nistir8547ipdPQCTransition} specifies a phased deprecation schedule for vulnerable algorithms such as RSA and the Elliptic Curve Diffie-Hellman, including a prohibition on their use after 2035. Beyond the NIST transition plan, several international initiatives further underscore the urgency of migrating to post-quantum cryptography, including the European Union’s coordinated implementation roadmap\cite{EU_PQC_Roadmap_2025}, which outlines a phased, cross-sector transition strategy and highlights the practical challenges of deploying PQC in real-world software systems.

Substantial work has also been undertaken to facilitate migration to PQC. This includes a systematic literature review on migration to PQC by Näther et al. \cite{Näther2024MigratingPQCSLR}, studies on network adoption rates \cite{Sowa2024PQCAdoptionRate}, and the creation of PQC libraries for developers such as PQClean \cite{Kannwischer2022PQClean} and Liboqs \cite{Stebila2017PQCTLS}. However, while PQC research has made substantial strides in algorithmic performance and protocol integration, limited research has investigated whether general software developers can correctly and securely implement these algorithms without security knowledge or cryptographic expertise.

Previous studies demonstrate that the usability of cryptographic libraries has a major influence on the security outcomes of applications \cite{Fahl2012AndroidSSLsecurity, Acar2017UsabilityCryptographicAPIs, Wijayarathna2019GoogleAPI}. When documentation is poor or APIs are unintuitive, developers are prone to misusing standard tools like Secure Sockets Layer (SSL), authentication mechanisms, and symmetric-key encryption. These usability issues often result in misuse in the application, such as the use of unsafe default settings or incorrect parameter configuration in classical cryptographic algorithms. To address these issues, Green \cite{Green2016UsableSecAPI} and Schmüser \cite{Schmüser2025CrypAPIDesign} have emphasized the importance of human-centered cryptography. However, at the time of writing, no study has empirically analyzed the usability of PQC APIs from a developer experience perspective.

Consequently, there is an urgent need for a systematic examination of the usability of the PQC API. Enhancing usability is critical for minimizing implementation errors of applications and promoting secure software development practices.

To address this gap, this research employed a moderated remote usability testing protocol facilitated via video conferencing software (specifically, Microsoft Teams). This format allowed participants to operate within their natural development environments while sharing their screens for real-time observation. Using the Cognitive Dimensions Framework (CDF) \cite{Wijayarathna2019CDFAPI, Wijayarathna2021methodologyCD} as an analytical lens, the study evaluated the interaction patterns of a diverse group of developers. The study compared two accessible APIs that represent distinct architectural models. Participants were tasked with implementing PQC algorithms in a simulated client-server environment, followed by a mixed-methods analysis to isolate the root causes of usability friction. 

This study examines the behavior of non-specialized developers and the experience of implementation under minimal manual intervention. It does not evaluate the security guarantees, cryptographic correctness, operational posture, or maturity of the deployment of the APIs examined. All findings relate solely to the experience of the developers and the usability of the APIs under experimental conditions. 

Specifically, this study seeks to explore key aspects of developer interaction with Post-Quantum Cryptography APIs by addressing the following research questions:
\begin{itemize} 
    \item [\textbf{RQ1}:] What common misuses and usability barriers do developers encounter when utilizing PQC APIs?
    \item [\textbf{RQ2}:] How do the integration challenges differ between endpoint-based PQC APIs and local library PQC API?   
    \item [\textbf{RQ3}:] What usability and guidance improvements are needed to support secure implementation?
\end{itemize}

The rest of the paper is outlined as follows: Section ~\ref{sec:literature_review}, ``Literature Review'', where we analyze previous studies relevant to our research questions. Section ~\ref{sec:methodology}, ``Research Methodology'' details our data collection and analysis methods. Section \ref{sec:Results}, ``Results'', where findings are presented with the aid of graphs and tables. Section ~\ref{sec:discussion}, ``Discussion and Evaluation'', which interprets and discusses the findings. In the concluding Section ~\ref{sec:conclusion}, ``Conclusion'', we summarize the insights gained from the experiment and suggest ways they could enhance future research.

 

\section{Literature Review}
\label{sec:literature_review}
This literature review first establishes the urgent need to migrate to PQC. Then it examines the challenges involved in this migration and identifies the stakeholders responsible for addressing them. Then, the review investigates how usability and documentation problems in PQC APIs can lead to insecure implementation behavior during application development. Finally, it explores the evaluation methodologies employed in related work to identify these issues and improve API design, including documentation, code examples, abstraction, and the inputs and outputs of functions.

\subsection{Importance of migrating to PQC}
Information security is entering a period of major change because a powerful quantum computer could break the classical encryption systems that protect today's digital world. Specifically, Shor's algorithm \cite{Shor1994, Shor1997}, a quantum computing algorithm, is theoretically capable of solving integer factorization and discrete logarithm problems in polynomial time. This capability would render the most widely used public-key cryptography algorithms, such as Elliptic Curve Diffie-Hellman (ECDH), Elliptic Curve Menezes-Qu-Vanstone (ECMQV), and Rivest-Shamir-Adleman (RSA), completely insecure \cite{Bernstein2009PQC, nistir8547ipdPQCTransition}. Consequently, the NIST transition plan mandates that these algorithms will be deprecated by 2030 and fully disallowed after 2035 \cite{nistir8547ipdPQCTransition}.

This threat is made more urgent by HNDL attacks, where encrypted data are harvested today to be decrypted by a future quantum computer \cite{Singh2024HNDL, Ali2021AnalysisPQC}. This is no longer a purely theoretical concern, as active research is consistently reducing the practical resource requirements for such an attack. For instance, recent optimizations of the windowed arithmetic circuits required for Shor's algorithm \cite{Shor1994, Shor1997} have been shown to reduce the total gate count for factoring by up to 3.4\%, incrementally lowering the barrier to a practical attack on gate-based quantum computers \cite{Luongo2025quantumrsa}. 

To mitigate this threat, the NIST standardization of algorithms like ML-KEM (FIPS 203)\cite{NIST:FIPS:203} introduces a paradigm shift based on the `KEM/DEM' framework \cite{Cramer2003KEM, shoup2001KEM}. In this hybrid approach, the KEM asymmetrically derives a shared secret, which is then utilized by a Data Encryption Mechanism (DEM) for symmetric message encryption. Consequently, the integration of these primitives shifts the burden from understanding mathematical theory to mastering practical software implementation \cite{Cherkaoui2024pqctransition}.



According to Näther et al. \cite{Näther2024MigratingPQCSLR}, this migration involves four key phases: diagnosis, planning, execution, and maintenance. While various roles such as Migration Managers and Security Experts are involved, the software developer emerges as the pivotal figure, holding primary responsibility for the critical execution and maintenance phases.

However, the central role of the developer is complicated by the intersection of limited domain expertise and the inherent complexity of the PQC algorithms themselves \cite{Näther2024MigratingPQCSLR}. Implementing PQC is a non-trivial task; as detailed in NIST guidelines SP.800.227 and FIPS 203 \cite{NIST.SP.800-227, NIST:FIPS:203}, a secure implementation must account for numerous sophisticated factors. These include managing secure handshake protocols, ensuring cryptographically secure random key generation, handling lattice set problems, and preventing side-channel attacks. Although addressing these factors is essential for compliance, this increased complexity often negatively impacts usability \cite{Laura2013usabilityanobstacle}. This tension between robust security and API usability remains an open research problem \cite{Burns2012SystematicMappingAPIusability, Murphy2018APIDesignersChallenges}, creating substantial friction for developers tasked with securing the next generation of digital systems.

\subsection{Usability Issues of Security API}
Historically, cryptographic APIs that exhibit limited usability have demonstrated substantial vulnerability to misuse by developers, resulting in major security flaws in applications \cite{Fahl2012AndroidSSLsecurity, Acar2017UsabilityCryptographicAPIs, Wijayarathna2019GoogleAPI}. This underlines the urgent requirement for empirical evaluation of the developer experience with emerging PQC APIs. Without usability evaluation, there will be risks of repeating past mistakes, where poor usability undermines the very security the migration aims to provide \cite{Stylos2007APIParametersObject, Dekel2009ImprovingAPIDocumentationUsability, Farooq2010APIusabilityPeerReviews, Clarke2004APIUsability, Myers2016APIUsability}.

Although the mathematical integrity of the new PQC standards is critical, history has demonstrated that mathematical security alone is insufficient to ensure real-world security \cite{Boyes2014heartbleedgotofail, Durumeric2014Heartbleed}. The cryptographic community has repeatedly learned this lesson from the failures of the classical cryptography API \cite{Boyes2014heartbleedgotofail}. The major vulnerabilities, such as the ``Heartbleed'' bug in OpenSSL \cite{Durumeric2014Heartbleed} or the Apple ``goto fail'' vulnerability \cite{Boyes2014heartbleedgotofail}, were not failures of the cryptographic algorithms themselves, but rather API implementation failures caused by developer error, complex code, and poor documentation \cite{Boyes2014heartbleedgotofail}.

Research in the usable security domain has consistently demonstrated that if an API is difficult to use, developers will make critical security errors in applications \cite{Green2016UsableSecAPI, Acar2017UsabilityCryptographicAPIs, Wijayarathna2019GoogleAPI}. Developers, who are often not security experts, may misconfigure parameters, mishandle sensitive data like secret keys, or fail to implement necessary procedures such as error handling or signature verification \cite{Green2016UsableSecAPI}.

However, these issues do not emerge as isolated single incidents; rather, they reflect recurring and well-documented misuse patterns. Frequent errors include the selection of inappropriate parameters, such as insecure modes such as ECB \cite{Egele2013CryptMisuseAndroid, Nadi2016JavaCrypAPI} and improper key handling practices, including hardcoding secrets or relying on unsafe default configurations \cite{Acar2017UsabilityCryptographicAPIs}. Developers also commonly omit essential security steps, such as certificate verification \cite{Georgiev2012SSLVulnerabilities, Fahl2012AndroidSSLsecurity}. For example, confusion and ambiguity in API usage have contributed to the widespread disabling of SSL / TLS verification in mobile applications. These failures often arise from inadequate documentation, the absence of robust secure-by-example guidance, an increased cognitive burden during implementation, and a reliance on insecure code snippets found through online forums and repositories \cite{Green2016UsableSecAPI, GorskiLoIacono2016UsabilitySecurityAPI, Myers2016APIUsability, Acar2016ImpactInformationSources}.

As the cryptography field advances toward PQC, these established misuse patterns pose even more risks. PQC introduces additional complexity that may further strain developer comprehension, including large key sizes, a novel key exchange model, inconsistent terminology, and the need for crypto-agility \cite{ott2019PQCChallengesAgility}. The need to implement secure hybrid schemes that combine classic encryption with PQC also introduces new risks of broken mechanisms or incorrect encapsulation logic \cite{Crockett2019PQChybridTLSSSH, Bindel2019HybridKEM}.

Furthermore, existing research highlights how even well-designed algorithms can fail in practice \cite{Green2016UsableSecAPI, Wijayarathna2019CDFAPI}, most usability studies have focused on traditional libraries like OpenSSL \cite{Ukrop2018OpenSSLCertificate}, BouncyCastle \cite{Wijayarathna2018UsabilityBouncycastleHasing}, and various Java or Python APIs \cite{WIJAYARATHNA2019JavaAPI, Acar2017UsabilityCryptographicAPIs}. This situation creates a critical research gap, as there has been little to no empirical evaluation of post-quantum-specific APIs. Therefore, this research aims to fill this gap by applying established usability evaluation methods to PQC APIs to identify usability issues that could lead to security vulnerabilities in applications.

\subsection{Related Work}
\label{subsec:related_work}

Research into the usability of cryptographic APIs has evolved from casual observation to structured and empirical evaluation. Acar et al. \cite{Acar2017UsabilityCryptographicAPIs} conducted a landmark quantitative study involving 256 Python developers to compare five libraries (PyCrypto, M2Crypto, cryptography.io, Keyczar and PyNaCl). Their methodology combined a controlled experiment with functional analysis, revealing a complex relationship between API design and security. Although comprehensive documentation facilitated functional correctness, it often led to insecure implementations, whereas overly simple APIs caused functional failures. Crucially, their demographic analysis noted that general programming familiarity did not correlate with security success; rather, specific security knowledge was the determining factor.

Complementing this quantitative approach, Wijayarathna and Arachchilage argue that identifying the root causes of insecure utilization of API requires qualitative depth. Through systematic literature reviews \cite{Wijayarathna2018SLRmethodologyEvaluateUseableSecAPI, Wijayarathna2019CDFAPI}, they evaluated various methodologies--including heuristic evaluations and API walkthrough--and concluded that empirical user studies are essential to reveal real-world developer experiences (DevX). To standardize this analysis, they adapted Clarke's CDF \cite{Clarke2004APIUsability}, expanding it from 12 to 15 dimensions to specifically address security contexts \cite{Wijayarathna2017CDQuestionaire}.

Applying this adapted CDF methodology, Wijayarathna and Arachchilage further investigated specific usability flaws across multiple environments. In their evaluation of the BouncyCastle API, they utilized the think-aloud protocol to identify that low-level parameters in the \texttt{SCrypt.generate()} method confused non-experts \cite{Wijayarathna2018UsabilityBouncycastleHasing}. Similarly, their assessment of the Google Authentication API revealed that misleading abstraction levels forced developers to rely on insecure third-party code snippets \cite{Wijayarathna2019GoogleAPI}. Furthermore, their study of the Java Secure Socket Extension (JSSE) API linked low penetrability and uninformative error messages to vulnerable TLS implementations \cite{WIJAYARATHNA2019JavaAPI}. These studies collectively demonstrate that when secure APIs are difficult to learn, developers inevitably revert to simpler, less secure alternatives.

In the domain of PQC, usability challenges are compounded by new cryptographic primitives. Zeier et al. \cite{ZeierAPIUsabilityStatefulSignature2019} addressed the complexity of stateful hash-based signature schemes (e.g., XMSS) by proposing ``EasySigner'', a crypto-agile API designed to abstract state management. Although their user study demonstrated high functional success, it highlighted a ``transparency paradox'': the effective abstraction left participants unaware that they were using a stateful scheme. This lack of awareness poses a risk, as developers might inadvertently compromise keys through external actions such as virtual machine cloning, underscoring the need for evaluation methods that assess both API usability and developer awareness. 

\begin{figure}[t]
    \centering
    \includegraphics[width=1.0\columnwidth]{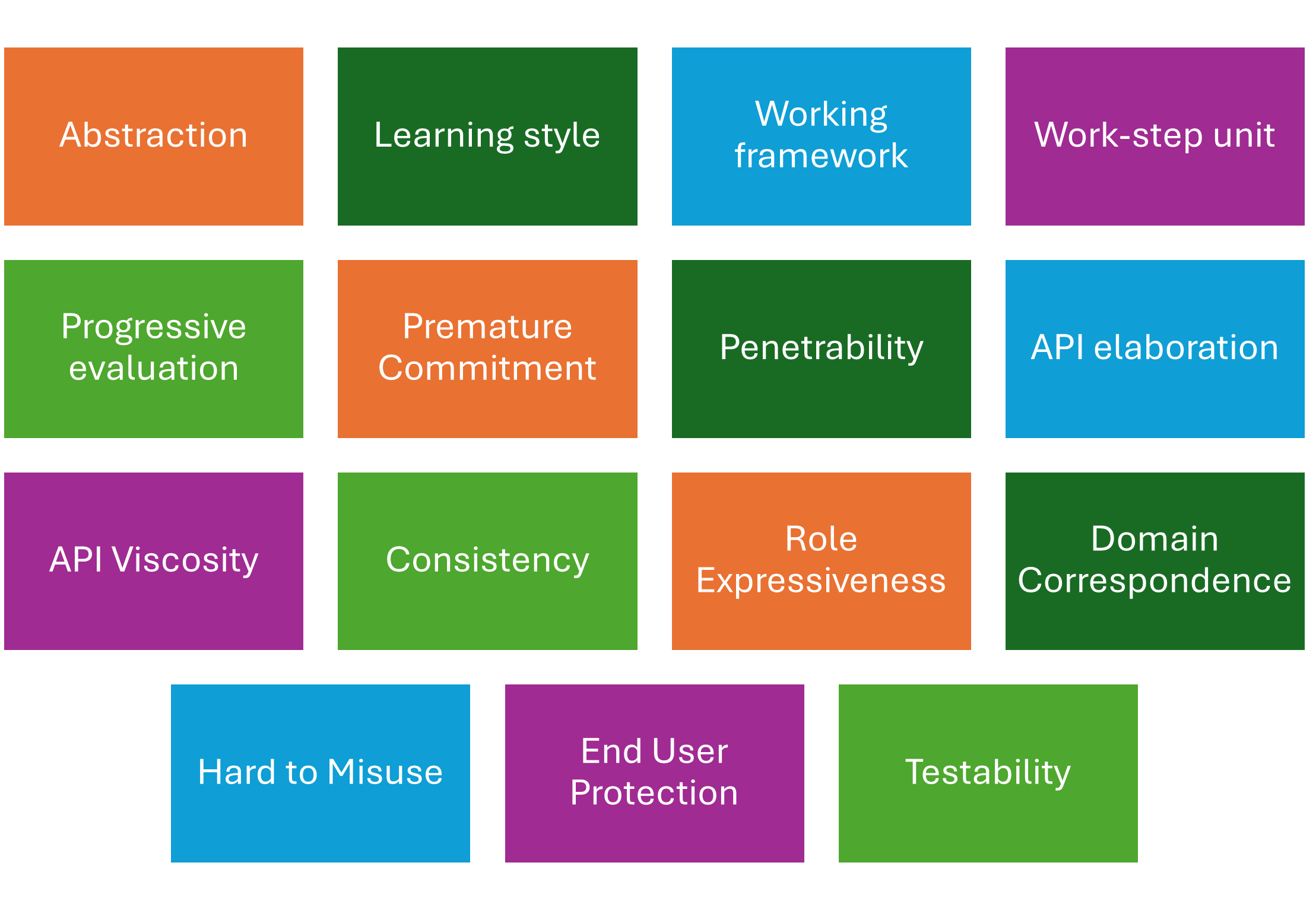}
    \caption{Cognitive Dimensions Framework.}
    \label{fig:15_cdf}
\end{figure}

\section{Research Methodology}
\label{sec:methodology}
This research aims to identify security vulnerabilities in applications that arise from usability issues that developers encounter when implementing PQC algorithms. As discussed in Section \ref{subsec:related_work}, we adopted an empirical user study methodology based on the CDF as summarized in \figurename~ \ref{fig:15_cdf} as modified by Wijayarathna and Arachchilage \cite{Wijayarathna2019CDFAPI}. This approach was selected to facilitate a detailed analysis of PQC APIs and to identify specific cognitive hurdles that hinder secure implementation. Aligning with observations by Acar et al. \cite{Acar2017UsabilityCryptographicAPIs} with respect to security experience and Näther et al.'s insights on PQC migration \cite{Näther2024MigratingPQCSLR}, we recruited a diverse set of participants ranging from software engineers and developers to IT students. This diversity allows the study to explore how varying levels of expertise influence security-relevant outcomes. Special emphasis is placed on developers without specialized cybersecurity training, as they are more likely to introduce implementation flaws in applications when interacting with complex or insufficiently supportive PQC APIs and libraries.

To ensure that the findings reflect real-world programming scenarios, we employed a moderated remote usability study design. This approach maximized ecological validity by allowing participants to use their familiar Integrated Development Environments (IDEs) and external resources, such as search engines and AI assistants (e.g., ChatGPT\footnote{https://chatgpt.com/}, Stack Overflow\footnote{https://stackoverflow.com/}). This was an intentional design choice to preserve ecological validity and to observe realistic developer–API interaction in natural development settings. To ensure consistency across sessions and to minimize environment-induced variability, all participants used the same programming language, identical task instructions, standardized skeleton code, and fixed versions of the evaluated PQC APIs. Environment requirements and dependencies were communicated in advance and verified at the start of each session, while live screen sharing and moderation allowed immediate resolution of environment-related issues. Consequently, observed differences are attributable to API usability and developer behavior rather than tooling or configuration discrepancies.

Data from pilot tests revealed that the cognitive load was excessive for a within-subjects approach. Therefore, the study was refined to a between-subjects model \cite{ALBERT201017PlanningUsability} where participants engaged with only one API. This design isolated the quantitative and qualitative analysis to a single interaction, removing the risk of learning bias.

The core assessment tasks were designed around the practical scenario of building a secure client-server communication protocol in compliance with NIST recommendations. Participants were required to implement two foundational PQC functions: a KEM, such as ML-KEM, and a DSA, such as ML-DSA. These tasks were presented sequentially to construct a layered security model: beginning with KEM to establish confidentiality, followed by symmetric-key encryption, and finally introducing DSA to prevent on-path attackers and ensure end-to-end integrity in the application.

The overall study procedure is illustrated in \figurename~ \ref{fig:study_procedure}. Before data collection, the study protocol was reviewed and approved by the university ethics committee. Following this approval, the process went on to recruit software engineers, developers, and IT students, who underwent a screening questionnaire to determine eligibility. Qualified participants received an informed consent form via email and were scheduled for a remote session. The session started with a briefing on general knowledge, followed by specific task guidelines( see appendix \ref{sec: task_guidelines}). The participants then executed the programming task using the ``think-aloud'' protocol \cite{Someren1994ThinkAloud}. Once the moderator verified the completion of the task, the session concluded with a post-task questionnaire adapted from Wijayarathna et al. \cite{Wijayarathna2019CDFAPI}. All data collected was then synthesized and mapped to the 15 Cognitive Dimensions to systematically categorize how API design features impact developer usability and software security.

\subsection{Programming Language and Chosen PQC APIs}
To minimize cognitive load during these tasks, the study utilized Python, selected for its readability and popularity in introductory programming \cite{Simon2018ProgrammingLanguage}. Participants were provided with a skeleton script containing helper functions for network socket communication, allowing them to focus strictly on the cryptographic implementation. Two PQC libraries--the endpoint-based PQ-Sandbox \cite{ExequantumDocs} and the local library QuantCrypt \cite{AabmetsQuantcryptWiki}--were selected solely based on public availability, PyPI accessibility, and completeness of public documentation. No API was selected due to institutional affiliation or endorsement. These criteria minimized installation barriers, serving as a methodological control to isolate usability issues inherent in the API design rather than environmental configuration. The PQ-Sandbox API \cite{ExequantumDocs} is a research prototype provided solely for academic evaluation. It is distinct from any production system and is intentionally simplified for experimentation. The API exposes only the cryptographic functions necessary for the study and should not be interpreted as a commercial or deployment-ready environment.

\subsection{Task design}
The task design was focused on the core functions of the PQC algorithms, namely KEM and DSA. We follow the NIST recommendations for secure KEM implementations \cite{NIST.SP.800-227}. These guidelines served as the standard for task design and implementation evaluation. Participants were given a set of materials to complete the task: two Python scripts containing skeleton code, API documentation, and a task instruction sheet.

The scenario placed the participant in the role of a software engineer at a financial technology company. They were assigned the task of developing a secure chat application project to prevent quantum computer attacks. The two scripts represented the server and client components of this application, which were assumed to operate on an insecure network. The participant's objective was to implement the PQ-Sandbox API and QuantCrypt API into this skeleton code to enable secure communication between the server and client on the insecure channel.

Task completion was measured against several goals: 
\begin{itemize}
\item \textbf{KEM}: The first task required participants to use ML-KEM to establish a shared secret between the server and the client. The fundamental security goal was to achieve confidentiality against a passive eavesdropper. By design, the KEM allows both parties to agree on a secret value without ever transmitting that secret directly across the network, thus protecting it from being intercepted. 

\item \textbf{Symmetric-key Encryption}: In the second task, participants were instructed to use the shared secret generated in Task 1 as a session key for a symmetric-key encryption algorithm such as the Advanced Encryption Standard (AES). This task demonstrates the ``hybrid encryption'' model, where the computationally expensive KEM is used only to establish the key, and the efficient symmetric-key cipher is used to protect the bulk of the application data.

\item \textbf{DSA for Handshake Authentication}: The third task directly addressed the on-path attackers vulnerability by introducing ML-DSA to provide server authentication. Participants had to modify the initial KEM handshake, requiring the server to use its long-term private key to sign a key component of the key exchange (such as the KEM ciphertext it generates). By verifying this signature with the server's trusted public key, the client can cryptographically confirm that it is communicating with the genuine server, not an impostor. This step adds the crucial security properties of authentication (proving the server's identity) and non-repudiation (providing undeniable proof that the server participated in that specific handshake).

\item \textbf{DSA for Message Exchange}: the final task was to implement message integrity and mutual authentication for the data-in-transit. Participants were asked to use ML-DSA on both the server and the client to sign every application message exchanged over the encrypted channel. This ensures that no message is altered by an attacker after it has been signed and confirms that both parties are who they claim to be throughout the entire session. Although Task 3 secured the handshake, it did not protect subsequent application data from being tampered with (even if it is encrypted). \end{itemize}

\subsection{Participant recruitment} 
Participants were recruited via posters with a QR code linking to a page outlining the study’s objectives and tasks, distributed through the researchers’ professional networks. Recruitment primarily attracted university students, staff, and professional developers from diverse sectors. Interested individuals completed a screening questionnaire assessing web API knowledge and programming proficiency to ensure foundational skills needed to prevent data noise. Eligible participants provided contact information to schedule sessions, while ineligible individuals were not asked for further details, protecting privacy and providing a clear rationale for ineligibility.

Eligible participants then received a Participant Information and Consent Form via email and, upon agreement, a researcher scheduled the experimental session based on their availability. A Microsoft Teams link confirmed the session. 

\subsection{The pilot study}
Before the main study, a preliminary pilot study was conducted with three participants. This small-scale trial was essential to ensure that the study ran smoothly and focused on testing the task design, rather than obtaining the final results. This process helped us determine the practicality, estimated duration, cost, and any unforeseen issues. These practical insights allowed us to refine our methods and proceed with the main study with greater confidence in our plan.

The pilot study revealed two critical challenges: technical setup hurdles and time constraints. First, to address difficulties with IDE configuration and data transfer, we refined the skeleton code by providing pre-built classes and included detailed environment requirements in the preparation email. This allowed participants to focus on core logic rather than troubleshooting. Second, because the average task duration exceeded 90 minutes, we assigned only one API per participant. Additionally, if a session surpassed two hours, we explicitly asked participants if they wished to continue. These strategies mitigated participant fatigue and preserved the quality of our research results.

\subsection{Study procedure}

\begin{figure}[t]
    \centering
    \includegraphics[width=1.0\columnwidth]{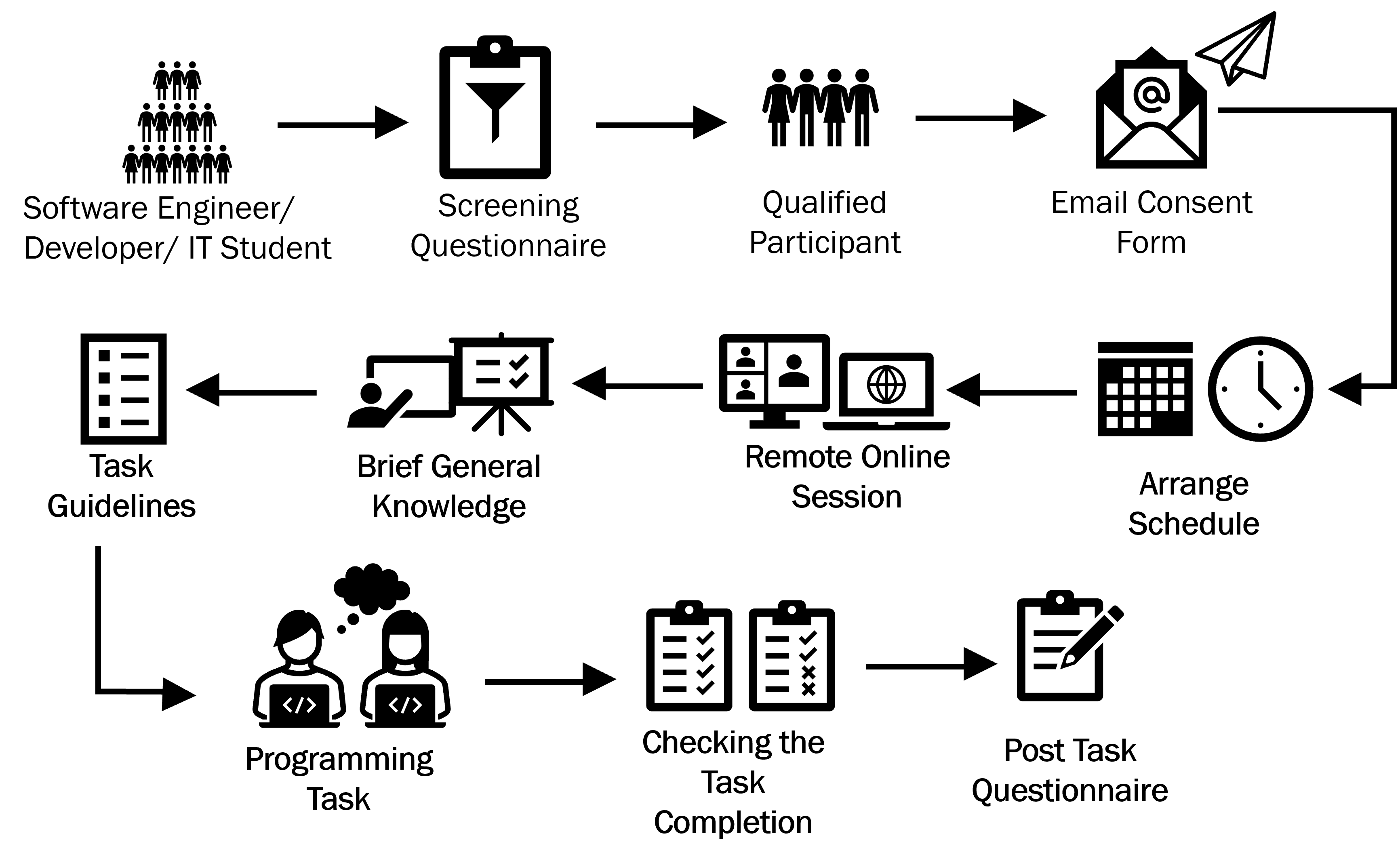}
    \caption{Study Procedure Diagram.}
    \label{fig:study_procedure}
\end{figure}

\begin{figure}[ht!]
    \centering
    \includegraphics[width=0.95\columnwidth]{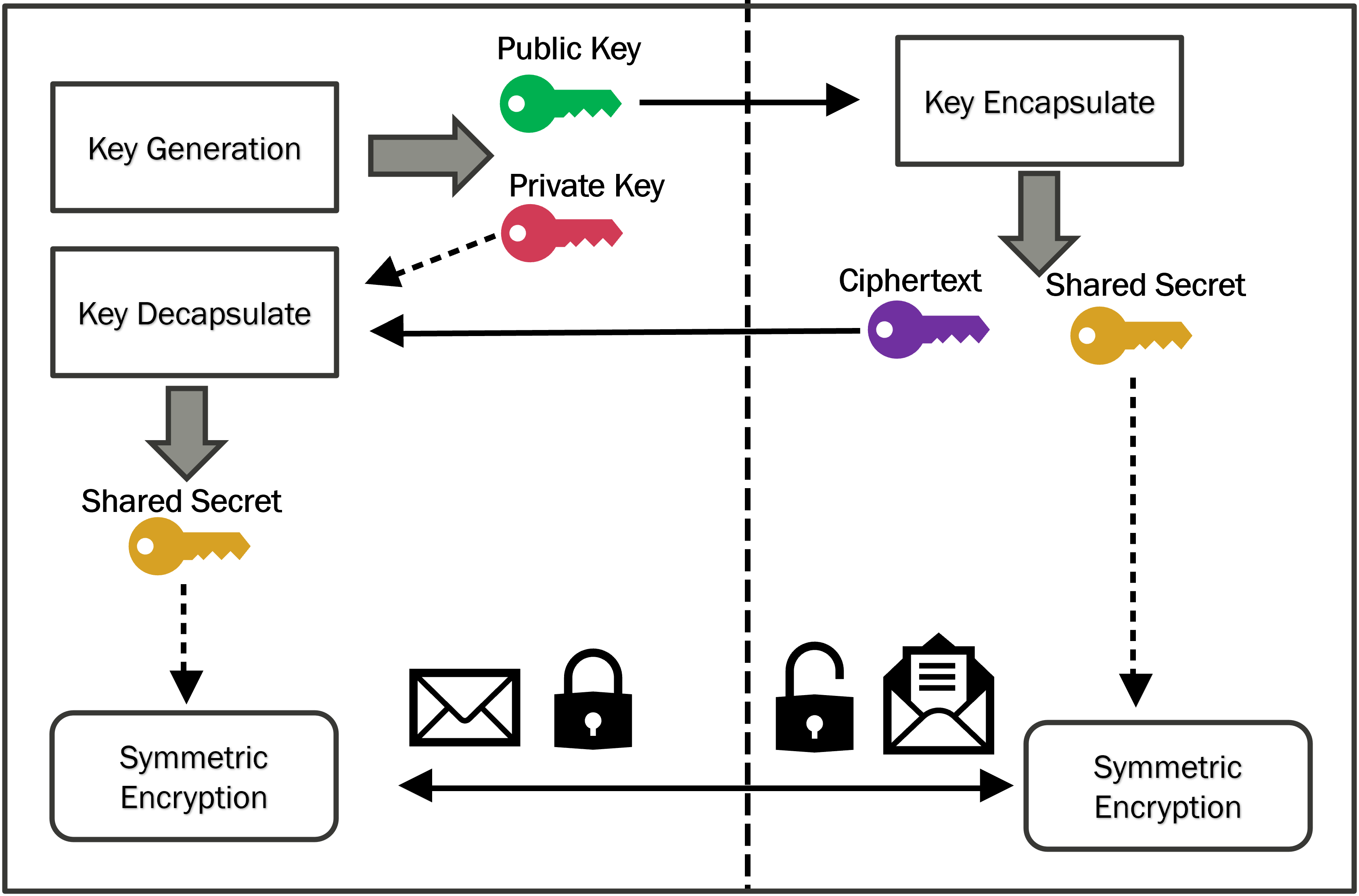}
    \caption{Key Encapsulation Mechanism and Symmetric Encryption Workflow.}
    \label{fig:kem_diagram}
\end{figure}
Following the pilot study and task refinements, we conducted the main usability study. The participants received a setup guide and a short video introducing the think-aloud method \cite{Someren1994ThinkAloud}.

The study workflow is shown in \figurename~ \ref{fig:study_procedure}. The protocol was approved by the University Ethics Committee before data collection began. The recruitment targeted software engineers, developers, lecturers, and IT students. Eligible participants, identified through a screening questionnaire, received an informed consent form and were scheduled for a remote session.

Each session opened with a brief introduction to the purpose of the study, background concepts, and task instructions. The participants then completed a programming task while verbalizing their thoughts using the think-aloud method \cite{Someren1994ThinkAloud}. After the moderator confirmed the completion of the task, participants completed a post-task questionnaire adapted from Wijayarathna et al. \cite{Wijayarathna2019CDFAPI}. All data were then mapped to the 15 Cognitive Dimensions to evaluate how API design influences usability and software security.

We conducted the study remotely, recording screen and audio to capture detailed navigation and reasoning processes often missed by surveys alone \cite{Ko2015empiricalhumanparticipant, ALBERT201017PlanningUsability}. Sessions began with a briefing on KEM and DSA concepts (e.g., \figurename~ \ref{fig:kem_diagram}) to ensure a common baseline before participants received the skeleton code. Participants performed tasks using a think-aloud protocol under passive moderation, followed by a post-study questionnaire. We prioritized participant comfort, explicitly allowing pauses or withdrawal at any time.




\subsection{Evaluation of PQC Usability}
The evaluation was based on the data collected, which included performance metrics, screen recordings, transcripts of the participants' ``thinking-aloud'' verbalization, and the post-task questionnaire. Quantitative analysis focused on performance metrics, including task completion rate, task completion time, and error rate of participants' implementations. The completion rate of the task was calculated as the percentage of participants who completed each task. Among the participants' implementations, the error rate was determined by counting the number of incorrect outputs, API misuse events, and runtime failures for each task. To compare performance differences between the two API groups (PQ-Sandbox and QuantCrypt), unpaired two-sample t-tests were used \cite{GraphPad_ttest_calc}. 

The qualitative data were analyzed using a thematic coding approach guided by the CDF \cite{Wijayarathna2021methodologyCD}. The first author performed the primary coding of all qualitative materials, including think-aloud transcripts and post-task questionnaire responses. A second author independently reviewed the coded data and the evolving codebook to validate interpretations and to identify potential inconsistencies. Any discrepancies were discussed and resolved through consensus, resulting in iterative refinement of code definitions and thematic boundaries. This validation process was employed to mitigate individual researcher bias and to strengthen analytical rigor. To preserve fidelity to participants’ perspectives, all quoted statements are reported verbatim, including original grammatical errors and typographical inconsistencies.

\section{Results}
\label{sec:Results}
\subsection{Participant Demographics}
\label{subsec:demographics}
A total of 16 participants were recruited for the study and randomly assigned to two independent, between-subjects groups: PQ-Sandbox ($N=8$) and QuantCrypt ($N=8$). One participant in the PQ-Sandbox group did not complete the post-task questionnaire. As summarized in Table \ref{tab:demographics}, the participants possessed varying levels of technical expertise. Regarding software development, the group included both beginners and professionals; while 7 participants were students or had no prior experience, the majority (9) were professional developers, with 6 possessing over three years of experience. Furthermore, all participants had at least some familiarity with Python, with half of the group ($N=8$) falling into the 1--3 year experience range. In terms of knowledge in the cybersecurity domain, the sample consisted predominantly of individuals with basic knowledge ($N=11$), while only three identified as experts and two reported no previous qualification or experience.

\begin{table}[t]
    \centering
    \caption{Participant Demographics.}
    \label{tab:demographics}
    \footnotesize 
    \renewcommand{\arraystretch}{1.2} 
    
    \begin{tabular}{p{2.4cm} p{4.0cm} c} 
        \toprule
        \textbf{Background} & \textbf{Category} & \textbf{N} \\
        \midrule
        
        \multirow{4}{=}{Software Development Experience (SDE)} 
          & No Experience & 3 \\
          & IT student (Entry Level Developer)  & 4 \\
          & $<$3 years (Beginner Developer) & 3 \\
          & $\ge$3 years (Expert Developer) & 6 \\
        \midrule
        
        \multirow{5}{=}{Python Experience (PE)} 
          & No experience & 0 \\
          & Less than 1 Year & 5 \\
          & 1 - 3 Years & 8 \\
          & 3 - 5 Years & 0 \\
          & $>$ 5 Years & 3 \\
        \midrule
        
        \multirow{3}{=}{Cybersecurity Experience (CE)} 
          & No & 2 \\
          & Basic Knowledge & 11 \\
          & Expert Cybersecurity & 3 \\
          
        \bottomrule
    \end{tabular}
\end{table}

\subsection{Task Performance}
\label{subsec:task_performance}

\begin{table}[ht]
    \centering
    \caption{Task Performance Time Metrics.}
    \label{tab:task_performance}
    \setlength{\tabcolsep}{3pt} 
    \small 
    \begin{tabular}{l c c c c}
        \toprule
        \textbf{API} & \textbf{Mean (min)} & \textbf{SD (min)} & \textbf{n} & \textbf{Comp. Rate} \\
        \midrule
        \multicolumn{5}{l}{\textbf{Task 1}} \\ 
        QuantCrypt & 39.38 & 15.52 & 8 & 100\% \\
        PQ-Sandbox & 65.38 & 18.02 & 8 & 100\% \\
        \midrule
        \multicolumn{5}{l}{\textbf{Task 2}} \\
        QuantCrypt & 41.43 & 6.88 & 7 & 87.5\% \\
        PQ-Sandbox & 25.43 & 7.72 & 7 & 87.5\% \\
        \midrule
        \multicolumn{5}{l}{\textbf{Task 3}} \\
        QuantCrypt & 25.60 & 5.77 & 5 & 62.5\% \\
        PQ-Sandbox & 23.25 & 6.02 & 4 & 50\% \\
        \midrule
        \multicolumn{5}{l}{\textbf{Task 4}} \\
        QuantCrypt & -- & -- & -- & 0\% \\
        PQ-Sandbox & -- & -- & -- & 0\% \\
        \bottomrule
    \end{tabular}
\end{table}

\begin{table}[htbp]
\centering
\caption{Participant Final Code Analysis.}
\label{tab:api_code_sec_analysis}
\setlength{\tabcolsep}{1.5pt} 
\scriptsize
\begin{tabular}{|c|l|c|c|c|c|c|c|c|}
\hline
\textbf{\raisebox{2em}{API}} & 
\textbf{\raisebox{2em}{ID}} & 
\rotatebox{90}{\textbf{KEM Decap Leak}} & 
\rotatebox{90}{\textbf{DSA Priv Leak}} & 
\rotatebox{90}{\textbf{Shared Sec Leak}} & 
\rotatebox{90}{\textbf{Missing Encap Check}} & 
\rotatebox{90}{\textbf{Missing Decap Check}} & 
\rotatebox{90}{\textbf{DSA Verify}} & 
\rotatebox{90}{\textbf{Key Not Destroyed}} \\ 
\hline

\multirow{9}{*}{\textbf{QC}} 
 & P1  & & & & \ding{55} & \ding{55} & \ding{51} & \ding{55} \\ \cline{2-9}
 & P2  & & & & \ding{55} & \ding{55} & & \ding{55} \\ \cline{2-9}
 & P5  & \ding{108} & & \ding{108} & \ding{55} & \ding{55} & \ding{55} & \ding{55} \\ \cline{2-9}
 & P6  & & & & \ding{55} & \ding{55} & \ding{55} & \ding{55} \\ \cline{2-9}
 & P9  & & & & \ding{55} & \ding{55} & \ding{55} & \ding{55} \\ \cline{2-9}
 & P11 & & & & \ding{55} & \ding{55} & & \ding{55} \\ \cline{2-9}
 & P13 & & & & \ding{55} & \ding{55} & \ding{55} & \ding{55} \\ \cline{2-9}
 & P16 & & & & \ding{55} & \ding{55} & \ding{55} & \ding{55} \\ \cline{2-9}
 & \textbf{Total} & \textbf{1} & \textbf{0} & \textbf{1} & \textbf{8} & \textbf{8} & \textbf{5} & \textbf{8} \\ \hline
 \hline

\multirow{9}{*}{\textbf{PQS}} 
 & P3  & & & & \ding{55} & \ding{55} & & \ding{55} \\ \cline{2-9}
 & P4  & & & \ding{108} & \ding{55} & \ding{55} & & \ding{55} \\ \cline{2-9}
 & P7  & & & \ding{108} & \ding{55} & \ding{55} & & \ding{55} \\ \cline{2-9}
 & P8  & & & & \ding{55} & \ding{55} & & \ding{55} \\ \cline{2-9}
 & P10 & & \ding{108} & & \ding{55} & \ding{55} & \ding{55} & \ding{55} \\ \cline{2-9}
 & P12 & \ding{108} & & & \ding{55} & \ding{55} & \ding{55} & \ding{55} \\ \cline{2-9}
 & P14 & & & \ding{108} & \ding{55} & \ding{55} & \ding{55} & \ding{55} \\ \cline{2-9}
 & P15 & \ding{108} & & & \ding{55} & \ding{55} & & \ding{55} \\ \cline{2-9}
 & \textbf{Total} & \textbf{2} & \textbf{1} & \textbf{3} & \textbf{8} & \textbf{8} & \textbf{3} & \textbf{8} \\ \hline

\end{tabular}

\vspace{1mm}
{\raggedright \tiny \textbf{Key:} \ding{108}=Leak Found, \ding{51}=Feature Implemented, \ding{55}=Missing Handling/Check \par}
\end{table}

Table \ref{tab:task_performance} summarizes the performance data from participants' initial API interactions. Results are grouped by PQC API and include the mean and standard deviation (in minutes), as well as the completion rate. While most participants successfully completed Task 1, failure rates increased for Tasks 2 and 3. Notably, no participants were able to complete Task 4 within the allotted time.

To evaluate performance on Task 1, an analysis was conducted to compare the average completion time (in minutes) between the QuantCrypt and PQ-Sandbox groups. The results revealed a clear and statistically significant difference, indicating that the QuantCrypt group was substantially faster. Participants using QuantCrypt finished the task in an average of 39.37 minutes, while the PQ-Sandbox group took considerably longer, averaging 65.38 minutes--a difference of 26 minutes. This conclusion is supported by an unpaired t-test ($t(14) = 3.09$, $p = 0.0079$), which shows that the probability that this large difference occurs by random chance is very low. Furthermore, we are 95\% confident that the true average advantage for the QuantCrypt group is between 7.97 and 44.03 minutes. Because this confidence interval does not include zero, it confirms that the observed performance gap is a genuine finding and not a statistical fluke.

For Task 2, a similar unpaired t-test was used. It revealed a difference, but with the opposite result ($t(12) = 4.09$, $p = 0.0015$). In this second task, the participant in the PQ-Sandbox group was significantly faster, with a mean completion time of 25.43 minutes, compared to the participant in the QuantCrypt group, who took significantly longer, a mean time of 41.43 minutes. On average, the PQ-Sandbox group took 16 minutes less to complete the second task. The 95\% confidence interval for this difference (7.49 to 24.51 minutes) again does not contain zero, confirming that the slower time observed in the QuantCrypt group is statistically significant.

Finally, in the analysis on Task 3 completion time, we found no statistically significant differences between the two groups ($t(7) = 0.596$, $p = 0.5700$). Descriptively, the mean time for the QuantCrypt group (N=5) was 25.60 minutes, and the mean of the PQ-Sandbox group's ($N=4$) was 23.25 minutes. This small difference of 2.35 minutes is likely due to random chance, a conclusion supported by the 95\% confidence interval, which ranged from -6.98 to 11.68 minutes. As this interval contains 0, it confirms the lack of a statistically significant difference in the implementation of ML-DSA using both APIs.

Within this primary data set (see appendix \ref{sec: task_performance_data}), Table \ref{tab:api_code_sec_analysis} details the security vulnerabilities identified in the participants' implementations and reveals distinct error patterns of the applications between the two groups. A notable and concerning observation was the complete absence of defensive programming across all participants' implementations. In their applications, irrespective of the library employed, every participant failed to incorporate error handling for KEM encapsulation or decapsulation, and none implemented explicit destruction of unused cryptographic keys. Despite this uniform lack of hygiene, differences emerged in the exposure to critical data from insecure applications written by participants. Participants using PQ-Sandbox demonstrated a higher rate of writing insecure applications, including shared-secret exposure and mishandling of key material. In contrast, the QuantCrypt group resulted in fewer data exposures in their applications, with only two leaks combined. However, this group struggled with the implementation logic, recording five instances of incorrect DSA verification handling in their applications compared to three among PQ-Sandbox users.

These outcomes highlight how introducing PQC to generalist developers remains inherently challenging. In the constrained study setting, documentation written for a cryptography-aware audience did not fully bridge the domain-knowledge gap, an issue mirrored across the broader PQC ecosystem.

\subsection{CDF Questionnaire}
Table \ref{tab:cd_mapping} presents the analysis of the CDF questionnaire results, mapping problematic dimensions to their corresponding themes. Dimensions for which participants reported no difficulties are listed but not mapped to a specific theme, as they did not contribute to the identified usability barriers.

\begin{table}[htbp]
    \centering
    \caption{Mapping of Cognitive Dimensions Problems to Identified Themes.}
    \label{tab:cd_mapping}
    \footnotesize 
    \renewcommand{\arraystretch}{1.2} 
    \setlength{\tabcolsep}{3pt} 
    
    \begin{tabularx}{\linewidth}{@{} l >{\RaggedRight\arraybackslash}X @{}}
        \toprule
        \textbf{Cognitive Dimension} & \textbf{Identified Theme} \\
        \midrule
        
        The Abstraction Level & API Complexity and Granularity \\
        \midrule
        
        Learning Style & \multirow{2}{=}{Documentation Deficiencies and Learning Barriers} \\
        Penetrability & \\
        \midrule
        
        The Working Framework & \multirow{3}{=}{Sequencing and Flow Dependency} \\
        The Work-step Unit & \\
        Premature Commitment & \\
        \midrule
        
        API Elaboration & \multirow{3}{=}{Naming, Data Handling, and Consistency Issues} \\
        Consistency & \\
        Role Expressiveness & \\
        \midrule
        
        Domain Correspondence & Importance of Prior Security Knowledge \\
        \midrule
        
        Error Proneness & \multirow{3}{=}{Security Dependence and Lack of Testability Guidance} \\
        End-user Protection & \\
        Testability & \\
        \midrule
        
        Progressive Evaluation & \textit{No Problem Found} \\
        API Viscosity & \textit{No Problem Found} \\
        
        \bottomrule
    \end{tabularx}
\end{table}

\subsubsection{API Complexity and Granularity}
Participants generally perceived the API as complex due to its low-level granularity, forcing them to manually combine separate cryptographic components. A majority (71\%) reported that multiple classes were needed to implement core functionality, contradicting the expectations of 59\% who anticipated a single-class solution. P8 (CE - Expert, SDE - Beginner) highlighted this discrepancy, noting that they \textit{``didn't expect that I would need to use and integrate multiple classes''} but rather assumed \textit{``a single entry point''} would handle the process. Instead, developers had to manually assemble the KEM, Key Derivation Function (KDF), and symmetric-key cipher. Although P8 felt that this assembly was achievable with structured guidance, P15(SDE - Expert) cautioned that \textit{``stitching them together... requires careful handling''} of endpoints and parameters, even if the modular roles were clear.

This requirement for manual assembly contributed to the consensus that the abstraction level was too low, exposing developers to \textit{``too many nuts and bolts''}. P5 (SDE - Expert) argued that developers expect to \textit{``minimize friction by just worrying about the relevant data''} rather than navigating internal cryptography jargon, while P1 (SDE - Expert) noted that manual key handling differed from typical key exchange mechanisms. Consequently, the volume of code required became a major friction point. Describing the workload for simple tasks as \textit{``unsustainable''}, P5 emphasized that code is a \textit{``liability in the long run''}, and that a developer typically expects to execute a task in \textit{``one liner or 2--3 lines max''} without managing the underlying interworking.

\subsubsection{Documentation Deficiencies and Learning Barriers}
Feedback highlighted hurdles in the learning process, particularly with respect to the depth and clarity of the documentation. Nearly half of the participants (44\%) felt that there was insufficient information, with P5 noting that code samples were often \textit{``unclear [or] misleading''} and P13 (SDE - Entry Level) finding the material unsuitable for those with \textit{``limited experience''}. This confusion was compounded by a lack of context; although the documentation detailed isolated function calls, it failed to illustrate the relationships between them. P2 (PE - $< 1$ year) criticized this approach, observing that, despite being extensive, the documentation lacked a \textit{``clear picture overview''} of the core mechanism required to get the system running.

The presentation style further alienated developers by relying heavily on academic terminology. P5 remarked that the \textit{``too many jargons''} meant a standard developer would struggle with half the content, leading P12 (CE - Expert) to suggest that \textit{``visual aids and a reduction in technical security jargon''} would make the concepts more accessible. Faced with these barriers, 63\% of the participants relied on copying code and \textit{``trial and error''} to understand the API. As P8 explained, the strict sequence of operations was not initially obvious, forcing them to derive the correct data flow by \textit{``ensuring outputs matched expectations''} rather than through clear instruction.

\subsubsection{Importance of Prior Security Knowledge}
The participants overwhelmingly agreed that the existing security knowledge was a critical factor in mitigating the difficulty of the API, with 88\% stating that prior experience would have facilitated the process. This created a distinct divide in user experience based on background. While P1 noted that the library is \textit{``hard to use... without prior knowledge''} due to the documentation lacking a clear flow overview, those with a foundational understanding fared considerably better. P16 (CE - Basic) reported that familiarity with concepts like public/private keys provided a \textit{``clearer picture of how the encryption and key exchange process actually works''}, preventing the disorientation felt by novices.

Specific technical competencies were often required to bridge the gap between documentation and the task. Experienced participants like P8 cited the need for \textit{``hands-on knowledge of ML-KEM and ML-DSA''} alongside general API integration skills. Consequently, the reliance on such specialized domains led some to perceive a mismatch in the intended audience. P6, identifying as a frontend developer, argued that the API was \textit{``mainly targeted for someone who has some experience in security''}, suggesting that without this specific context, the implementation barrier remains high for generalist software engineers.

\subsubsection{Sequencing and Flow Dependency}
The API workflow was defined by a rigid sequential structure, with 88\% of the participants reporting that the system forced them to think ahead and prioritize specific decisions. This workflow dictated a strict execution order--key generation $\rightarrow$ encapsulation / decapsulation $\rightarrow$ deriving shared secret $\rightarrow$ encryption / signing--which required developers to preemptively plan their architecture. P15 emphasized the cognitive load of this planning, stating that success required understanding \textit{``who generates keys, when to send the public key, and when to encapsulate/decapsulate''} before implementation. However, this dependency chain was not immediately intuitive; 63\% of participants admitted to identifying these necessary advanced decisions through trial and error, often struggling to determine the correct operational order for complex components like encapsulation and encryption.

\subsubsection{Naming, Data Handling, and Consistency Issues}
Participants encountered friction with respect to ambiguous naming conventions and unclear data requirements. The use of abbreviations like ``\texttt{pk}'', ``\texttt{sk}'', and ``\texttt{data}'' was criticized as non-intuitive, with P10 (SDE - Expert) arguing that these were \textit{``confusing''} and poor conventions. Similarly, P5 felt that method names such as ``\texttt{keygen}'' sounded unprofessional--likening it to \textit{``pirated software''}--and suggested more standard alternatives like ``\texttt{generateKey()}''. This lack of clarity extended to data definition; P15 noted that functions such as \texttt{encrypt\_text} did not specify necessary input properties, while P16 reported that the data types for the parameters and the return values were generally \textit{``difficult to find''}.

Beyond terminology, manual data manipulation and functional consistency presented challenges. P16 highlighted a specific hurdle where the API produced a 32-byte key while the \texttt{Krypton} class required a 64-byte key, forcing them to \textit{``look for external resources''} to handle the conversion. Users also reported initial confusion regarding the similarity of certain functions. P2 noted that \textit{``some functions looked similar at first because their names were not very clear''}, requiring documentation checks to distinguish them. However, P8 offered a counterpoint, observing that while KEM and KDF had similar goals, they were \textit{``clearly separated by role''}--one for exchange and one for shaping secrets--which helped clarify the distinction within the workflow.

\subsubsection{Security Dependence and Lack of Testability Guidance}
Participants acknowledged that security was highly dependent on their correct implementation, yet they often lacked the necessary guidance to verify it. A majority (80\%) understood that end-user security depended on both the API's guarantees and their own code, specifically regarding key handling and sequencing. P8 articulated this distinction, noting that while the API provided primitives like ML-KEM, the \textit{``actual security outcome''} relied on the developer ensuring \textit{``correct sequencing, validation, and correct handling of keys''} during the key exchange. However, maintaining this security was complicated by opaque error reporting. P8 noted that most issues, such as decapsulation failures, had to be handled at the program level, while P5 expressed frustration that generic error messages like \texttt{CipherVerifyError} provided \textit{``not enough info as to `why'\,''}, forcing them to \textit{``ask AI to help debug''} rather than relying on API feedback.

Despite these risks, the majority of participants (67\%) did not test the security of their applications, citing time constraints and lack of instructional support. P2 admitted that they skipped testing because they were \textit{``wasn't sure how to do it properly''} and required specific examples. This uncertainty reflected a broader gap in documentation; 57\% were unsure if testing guidance existed and 36\% stated that it was absent. P8 criticized the lack of context regarding how the algorithms matched \textit{``NIST PQC standards''} or regulatory requirements, while P16 noted that the documentation failed to explain the \textit{``key exchange process and why a 64-byte symmetric key was required''}, forcing developers to rely on external sources to validate their security posture.

\section{Discussion}
\label{sec:discussion}
\subsection{RQ1: What common misuses and usability barriers do developers encounter when utilizing PQC APIs?}
The evaluation revealed recurring usability patterns that align with previous research on classical cryptographic APIs. These patterns highlight how introducing PQC to generalist developers without expert onboarding or contextual guidance creates predictable friction. The findings reflect the natural gap between rapidly evolving PQC standards and the mental models of developers who are encountering these primitives for the first time.

Many participants mishandled key material, for example, sending decapsulation keys or sharing secrets across the network. These behaviors stemmed from a missing conceptual understanding of typical KEM / DSA workflows when working without onboarding, as well as documentation gaps. In particular, many participants assumed that any value output by a function must be transmitted, a common behavior documented in past usability research. Clarifying examples in future documentation, especially emphasizing that shared secrets never leave local memory, would reduce such misunderstandings.

Both APIs used established cryptographic abbreviations such as (\texttt{pk} / \texttt{sk}). Although these are standard within the cryptography community, several participants unfamiliar with such conventions found them difficult to interpret. This challenge was amplified by the fact that NIST released new educational guidance, such as the Encapsulation / Decapsulation Key terminology in SP 800-227--after the API documentation used in this study had already been written. As a result, the documentation and the newer NIST teaching examples diverged slightly in terminology, leaving participants without the contextual anchors they would normally rely on in a production setting. This mismatch reflects the evolution of the natural ecosystem and highlights the importance of aligning terminology across the PQC ecosystem as standards mature.

\begin{figure}[ht!]
    \centering
    \includegraphics[width=\linewidth]{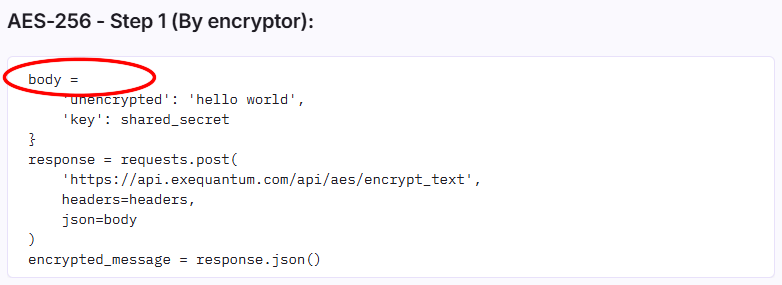}
    \caption{Mistake on PQ-Sandbox Documentation (Source: \cite{ExequantumDocs}).}
   \label{fig:typo_pqcaas}
    \vspace{0.5cm} 
    
    \includegraphics[width=\linewidth]{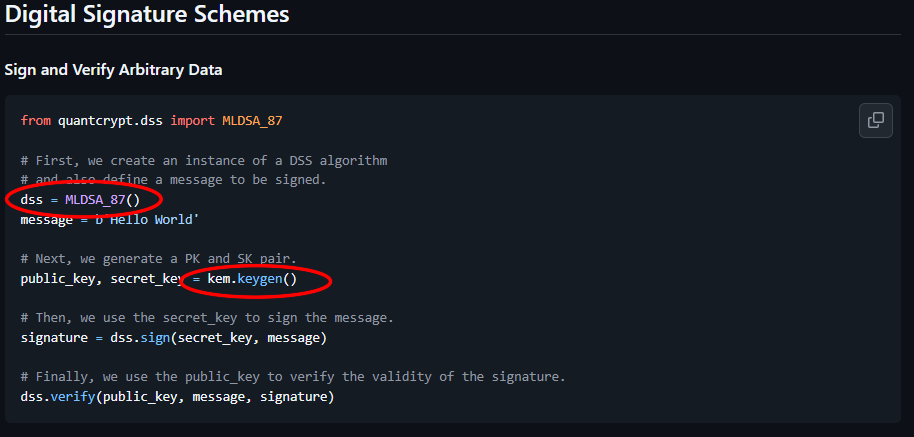}
    \caption{Mistake on Quantcrypt Documentation (Source: \cite{AabmetsQuantcryptWiki}).}
    \label{fig:typo_quantcrypt}
\end{figure}

Minor documentation inconsistencies (e.g., typos, missing brackets) were found in both APIs. Specifically, neither API provides comprehensive explanations of variable types or the nature of the outputs returned by their functions. This lack of detailed guidance limits non-specialized developers' ability to correctly interpret API behavior, increasing the potential for implementation errors and reinforcing usability-related security risks in the applications. There are also some typos, such as incorrect variable names (see \figurename~ \ref{fig:typo_quantcrypt})  or missing brackets (see \figurename~ \ref{fig:typo_pqcaas}) in the syntax. 
Although these seem like minor issues for experienced developers, for someone who lacks experience in programming, they might not know where the error is. Inconsistency in variable naming and error output also became a pain point for participants, as they were confused by changing terminology, and the error output was not explained in the documentation.

A further observation concerned the developer's handling of transient key material at the application layer. Participants did not destroy intermediate keys or ephemeral secrets, as recommended in NIST SP 800-227\cite{NIST.SP.800-227}. Neither API is designed to manage memory erasure, and Python's memory model does not provide built-in, high-assurance primitives to securely erase sensitive data; object lifetime and copies are managed by the interpreter and garbage collector, so this behavior was not a functionality flaw or non-compliance of either API with NIST's guideline. However, the unfamiliarity of PQC workflows may argue that documentation across the broader PQC ecosystem may benefit from clearer conceptual guidance (beyond the documentation on the functionality alone) on application-layer key-lifecycle practices, particularly for non-specialist developers.

The evaluation revealed testability issues in both APIs, as most participants did not write tests for their cryptographic applications. Although time constraints contributed to this behavior, but unclear documentation and lack of testing examples were also factors. Participants often overlooked error-handling mechanisms, highlighting the need for documentation to provide explicit guidance on best practices. These findings indicate an opportunity for the PQC ecosystem to improve support for verifying correctness when using novel cryptographic primitives.

\subsection{RQ2: How do the integration challenges differ between endpoint-based PQC APIs and local library PQC API?}

The results and analysis that informed the RQ2 reveal differences and clear trade-off between the QuantCrypt API and the PQ-Sandbox API design choices. Participant performance and perceived usability varied between the two APIs, reflecting the distinct cognitive and operational demands imposed by each. These variations are expected, given the fundamentally different design principles and levels of abstraction underlying the two APIs, which shape both the ease of use and the types of errors developers are likely to encounter.

The QuantCrypt API provided a direct KEM implementation, but concealed several underlying complexities. A major usability challenge emerged when participants attempted to integrate the shared secret generated by KEM with a symmetric-key cipher. Most participants did not recognize that the KEM output was 32 bytes and was required to process through a KDF to produce a 64-byte key suitable for subsequent AES encryption. This gap illustrates how low-level abstractions, while flexible, can impose substantial cognitive overhead and increase the risk of implementation errors in applications.

For context, ML-KEM outputs a 32-byte shared secret by design. QuantCrypt lets developers to apply their own Key Derivation Function (KDF) when a 64-byte key is needed for AES or similar ciphers. In contrast, the PQ-Sandbox prototype applies a KDF internally and returns a 64-byte symmetric key directly. This architectural difference explains much of the divergence in task completion times for Task 2.

Several participants misinterpreted example snippets because they lacked context on how the example differed from their task scenario. This reflects the limitations of the scope of the documentation under experimental conditions. Participants often did not know that the source of the problem was in their initial use of a random key instead of the KEM-generated shared secret. Resolving the issue required line-by-line backtracking to identify the origin of the error, increasing the cognitive load, and could lead to further developer errors. This confusion directly explained the reasons for the longer task completion times observed for Task 2 among the QuantCrypt group.

In contrast, the PQ-Sandbox API operates at a different level of abstraction for the KEM function. It integrates the KDF step directly within the KEM operation. As a result, the shared key produced by the KEM was already 64 bytes and could be used immediately for symmetric-key encryption, eliminating the confusion observed with the QuantCrypt API. Nevertheless, PQ-Sandbox introduced its own usability challenges, primarily related to its endpoint-based design. Participants were required to create custom objects or methods to access API functions and frequently struggled with implementing authentication headers and tokens to use the API, which proved confusing. This initial setup overhead contributed to slower completion times for Task 1 among PQ-Sandbox users compared to QuantCrypt. However, once participants became familiar with these procedures, the consistency and similarity of subsequent function calls facilitated the implementation of ML-DSA in Task 2, making the process more straightforward than for QuantCrypt users.

Both APIs leave the key rotation policy to the application layer. Developers unfamiliar with such practices may benefit from higher-level documentation guidance or examples. QuantCrypt exposed a KDF that could be leveraged to derive multiple keys from the same shared secret. With PQ-Sandbox, generating a new key required repeating the entire KEM process, a time-consuming procedure that could unintentionally lead to insecure utilizations in the application when lacking clear documentation or examples, such as reusing keys across multiple sessions. This design choice highlights a trade-off between abstraction convenience and the flexibility required for secure key management.

Furthermore, the design of the PQ-Sandbox, which prioritizes security and privacy, further ensures that the generated keys are not stored on the server and are instead output directly to the users. However, this approach leads to developers receiving multiple keys for different purposes during the DSA signing process, which confuses non-experts. The DSA signing step returns a key pair in the prototype environment. In the constrained-study setting, several participants were unfamiliar with how signature schemes typically separate long-term identity keys from ephemeral signing keys. Without explicit onboarding explaining these roles, some participants incorrectly assumed that all returned fields needed to be transmitted. The observed misunderstandings were attributable to gaps in participant familiarity with DSA workflows and the limited conceptual guidance provided in the study material.

\subsection{RQ3: What usability and guidance improvements are needed to support secure implementation?}
The research attempted to identify potential solutions to the observed issues through the third research question. The findings indicate that the security vulnerabilities uncovered in participants' implementations stem not merely from individual developer mistakes but from structural usability obscurities embedded within the architectures of APIs themselves. Addressing these shortcomings, therefore, requires a developer-centric design philosophy that guides non-expert users toward secure practices in the API designs and documentation.  Drawing on insights into mental models and workflows of developers, the proposed recommendations focus on improving usability and developer performance while ensuring that secure development practices are reinforced by default in the applications.


\subsubsection{Adopt a Secure-by-Default Design}  
The absence of implemented key destruction and error handling in the applications highlights that developers cannot be expected to understand that they need to manage such tasks manually, particularly when documentation offers no guidance or examples. In addition, the confusion surrounding different key lengths and the necessity of KDF usage highlights a broader need for improved API design. One sound solution is to offer additional high-level functions that encapsulate common cryptographic workflows. Such an approach will reduce developer misuse and cognitive load while still allowing experts to access low-level primitives for advanced users who require greater control.

\begin{itemize}
    \item \textbf{Automate Key Life Cycle}:       
    To reduce the risk of secret-key exposure in the application, we recommend that cryptographic APIs manage the full key life cycle, including timely memory cleanup. Ideally, this is achieved through native automation, for example, binding private keys to constructs like Python's ``with'' statement so that keys are wiped when they leave scope. If such mechanisms cannot be built in, such as due to compliance constraints (i.e., highly regulated environments), we recommend that the cryptographic APIs should still guide developers by integrating secure-memory libraries or, at a minimum, providing clear documentation and examples that demonstrate proper key-destruction procedures and explain their security rationale.
    
    \item \textbf{Provide High-Level, Secure-by-Default Functions:} We argue that cryptographic APIs should, by default, encapsulate complex security operations, particularly for developers without specialized cybersecurity expertise. A high-level function, such as \texttt{establish\_secure\_channel()}, should set up the full sequence of operations, including KEM, KDF, and handshake authentication internally, and these different cryptographic components should be used and assembled following the best security practice by default. The low-level functions should remain accessible, but explicitly marked for expert use, and should be used with caution. If such high-level functions are not built into the cryptographic API, then the documentation must compensate by providing complete, secure, and executable example workflows that demonstrate best practices and prevent misuse.
  
    \item \textbf{Enforce Error Handling}: Furthermore, cryptographic functions should not return boolean or status codes that developers can ignore. Failures in decapsulation or signature verification are security-critical and should raise explicit exceptions by default, forcing developers to handle them properly.
\end{itemize}

\subsubsection{Prioritize Developer-Centric Documentation}
Documentation has consistently emerged as a central usability barrier. It should be revised from a developer-oriented perspective rather than that of a security expert designing the API. Effective documentation is essential to help developers build secure applications while reducing the unnecessary cognitive burden.

\begin{itemize}
    \item \textbf{Create ``Flow'' Documentation:} The documentation should evolve from a function-by-function reference into a comprehensive developer guide. It should present high-level use cases and include diagrams that illustrate the complete client–server workflow from start to finish. Examples should align with trusted guidelines, such as NIST, and demonstrate correct usage, common pitfalls, and how to apply the code in real-world applications.
    
    \item \textbf{Provide Runnable, Secure-by-Default Examples:} The documentation should include complete ``copy-and-paste'' examples that demonstrate the full secure workflow, for example, KEM and DSA with correct error handling and secure key management. Such examples serve as practical best-practice models, a need emphasized by participants who noted that the provided skeleton code was essential for understanding how to proceed. These examples must also be contextualized, reflecting realistic scenarios, such as establishing a secure communication channel rather than presenting isolated function calls.
    
    \item \textbf{Use Clear and Standardized Terminology:}  APIs and their documentation should employ intuitive, self-explanatory, and unambiguous names for functions and variables to improve readability and help developers understand expected behaviors. To minimize confusion, APIs should adopt the NIST recommended terminology, using ``encapsulation key'' and ``decapsulation key'' instead of generic ``public key'' and ``secret key'', and avoid unclear abbreviations such as ``\texttt{sk}'', ``\texttt{pk}'' and ``\texttt{enc}'', which increase cognitive load. Recognizing that not all developers are cybersecurity experts, these clear names should be accompanied by explicit explanations and descriptions in the documentation. Following NIST naming conventions would standardize terminology across PQC APIs and make them more accessible to general developers.
    
\end{itemize}

\section{Limitations}
This study examined general developer behavior under limited documentation and limited onboarding, reflecting a low-support environment rather than real-world operational deployments or cryptographic experts' deployment. The findings are based on a small sample of participants, which limits quantitative generalizability but provides strong qualitative insight into early-stage usability patterns. Additionally, the study focuses on the developer experience rather than the security, performance, or production readiness of the assessed APIs.

\section{Conclusions}
\label{sec:conclusion}
This study examined the usability of two Post-Quantum Cryptography (PQC) APIs, QuantCrypt, a local library, and PQ-Sandbox, an endpoint-based API, among developers without deep cybersecurity expertise. Both APIs posed challenges in integrating PQC algorithms, with design differences affecting developer interaction. Usability limitations contributed to implementation errors, showing that strong mathematical security alone is insufficient without developer-friendly tools.

Results revealed a performance-usability trade-off, while highlighting mismatches between API design assumptions and developer expectations. Participants struggled with jargon-heavy documentation, domain knowledge gaps, and abstraction challenges requiring trial-and-error assembly of cryptographic components. These findings emphasize that simply providing PQC algorithms in an API is insufficient for application security; a developer-centric focus on usability is critical to prevent insecure implementations of applications.

Based on these insights, we recommend that PQC API designers should prioritize usability, providing secure-by-default, high-level functions, NIST-compliant naming, and clear, illustrated documentation with executable examples. Future work should involve larger studies and the development of a high-usability PQC API addressing the failures identified here, ensuring that cryptographic tools effectively shield developers from the application vulnerabilities they aim to prevent.



\section*{Acknowledgment}
The authors thank Associate Professor Nalin Arachchilage for his supervision and guidance. We gratefully acknowledge Samuel Tseitkin and ExeQuantum for their technical support with the APIs and Sandbox environment. We also extend our appreciation to the study participants for their time and valuable contributions. This work was supported by the Australia Awards Scholarship (AAS).



%


\bibliographystyle{IEEEtran}
\bibliography{References}

\appendices
\section{Example of Mistake on Documentation}

\section{Task Guidelines}
\label{sec: task_guidelines}
The purpose of these tasks is to evaluate the usability issue that might arise when developer implement Post-Quantum Cryptography (PQC) algorithm on software using PQC APIs. Based on the implementation of API, it is categorised by two types of APIs: Web-based PQC API (accessed via HTTP requests) and traditional PQC API (accessed via a local library/package). You will be performing real-world–inspired programming tasks that simulate integrating PQC API into a simple software application.

\subsection{Introduction (2 minutes):}
\begin{itemize}
    \item ``Thank you for participating in this usability test. We're evaluating the usability of new cryptographic algorithms, specifically Post-Quantum Cryptography (PQC), which includes the Key Encapsulation Mechanism (KEM) and Digital Signature Algorithm (DSA). These algorithms are designed to be secure against future quantum computers, but there might be some usability issues for developers when they implement them using an API.''
    \item ``Your task is to implement a simplified secure communication channel between Server and Client, using these algorithms. Please think aloud as you work, explaining your steps and any challenges you encounter.''
    \item ``You are free to consult documentation if needed, but we'd like to see how intuitive the APIs are initially. There are no right or wrong answers; we are testing the technology, not you.''
    \item ``Do you have any questions before we begin?''
\end{itemize}

\subsection{Goal and Framing (2 minutes):}
\begin{itemize}
    \item ``Your objective is to Integrate the PQC API into a provided skeleton program to send a secure message between two server and client.''
    \item ``Imagine you are a software engineer at a mid-sized financial technology company. Your team is preparing for the future where traditional cryptography may no longer be secure against quantum computers. To ensure customer data remains protected, your manager has asked you to prototype the use of Post-Quantum Cryptography (PQC) algorithm on the software using PQC API. You need to generate the shared secret (using Key Encapsulation Mechanism)  between the server and client. This shared secret will be used as key on symmetric encryption, so server and client will exchange information in encrypted way.''
\end{itemize}

\subsection{Instruction (3 minutes):}
\begin{itemize}
    \item ``You will be given two python script, server.py and client.py. you need to open these scripts on your IDE and you need to runt the server first before running the client''
    \item ``You are free to use any tools that you usually use when developing software. But please share the screen when you are doing this and think out loud when you use this tools''
    \item ``While performing the tasks, you need to talk aloud what you are thinking, so it can be recorded with screen recording.''
    \item ``Before starting, make sure your microphone is unmute and choose share the whole screen. If you work using  two or more monitor please make sure what you work only on one screen so what you read and work could be captured on share screen.''
    \item ``Do you have any questions?''
\end{itemize}

\subsection{Task Explanation (1 minutes):}
\begin{itemize}
\item ``You will be given to four tasks''
\item ``First Key Encapsulation and Decapsulation Mechanism''
\item ``Second Symmetric encryption and decryption''
\item ``Third Digital Signature Algorithm for Handshake Authentication Protocol''
\item ``Fourth Digital Signature Algorithm for Message Exchange''
\end{itemize}

\subsection{Task 1: Key Encapsulation and Decapsulation (15-20 minutes):}
\begin{itemize}
    \item ``Now, let’s simulate Server and Client establishing a shared secret using ML-KEM.''
    \item ``Your first step is to make Client generate key pairs (Public and Private Key) for ML-KEM.''
    \item ``Then Client will have to send client public key to the server.''
    \item ``Server will receive client public key and use it to generate ciphertext and shared secret.''
    \item ``Server will send this ciphertext to Client.''
    \item ``Client should then use his ML-KEM private key to decapsulate the ciphertext and recover the shared secret.''
    \item ``Finally, please implement a check to verify that the shared secrets generated by Server and Client are identical.''
\end{itemize}

\subsection{Task 2: Symmetric Encryption (15-20 minutes):}
\begin{itemize}
    \item ``Now, after server and Client got shared secret, this shared secret need to be used for symmetric encryption.''
    \item ``Then Server will have to use this key to encrypt the message and send this message to client.''
    \item ``Client will receive server encrypted message and decrypt it.''
    \item ``Show the decrypted message on client.''
    \item ``Next Client will encrypt the message and send this encrypted message to server.''
    \item ``Server should decrypt the message from client and show it on server side.''
\end{itemize}

\subsection{Task 3: Digital Signature Algorithm and Verification for Handshake Authentication Protocol (15-20 minutes):}
\begin{itemize}
    \item ``Next, we'll focus on authentication using ML-DSA.''
    \item ``Now, let’s combine these pieces into a simplified secure handshake. Imagine Client initiating a secure communication with Server.''
    \item ``Client generates an ML-KEM key pair and sends its public key to Server.''
    \item ``Server then uses Client’s public key to encapsulate a shared secret and sends the resulting ciphertext.''
    \item ``Crucially, Server also signs the entire message its sends to Client (which includes the ML-KEM ciphertext and his own public key) using its ML-DSA private key.''
    \item ``Client receives this message (ciphertext, Server Public Key, and Server Signature). Client should verify Server's signature using Server public key, then if it is true client proceed the process to decapsulate the shared secret using its ML-KEM private key.''
    \item ``The goal is to have a functional handshake where Client has a shared secret with Server and can be confident that the message (including the encapsulated secret and Server's identity) came from Server.''
\end{itemize}

\subsection{Task 4: Digital Signature Algorithm and Verification for Message Exchange  (15-20 minutes):}
\begin{itemize}
    \item ``Next, we'll focus on authentication using ML-DSA.''
    \item ``Server has a message or document she wants to send to Client securely. Please implement a process where Server signs this message using its ML-DSA private key.''
    \item ``Client should then implement a way to verify Server’s signature using Server ML-DSA public key.''
\end{itemize}

\subsection{Post Task Questionnaire (20-30 minutes):}
\begin{itemize}
    \item ``After you finish all the task or finalized the experiment you could scan the qr code or click the link on this task guideline to fill the post task questionnaire.''
\end{itemize}

\section{Task Peformance Raw Data}
\label{sec: task_performance_data}
\begin{table}[htbp]
    \centering
    \caption{Participant Task Completion Time (minutes)}
    \label{tab:task_data}
    \setlength{\tabcolsep}{4pt} 
    \scriptsize
    \begin{tabular}{|c|l|c|c|c|c|}
        \hline
        \textbf{API} & \textbf{ID} & \textbf{T1} & \textbf{T2} & \textbf{T3} & \textbf{T4} \\
        \hline

        \multirow{8}{*}{\textbf{QC}} 
        & P1  & 23 & 43 & -- & -- \\ \cline{2-6}
        & P2  & 27 & 38 & -- & -- \\ \cline{2-6}
        & P5  & 40 & 42 & 31 & -- \\ \cline{2-6}
        & P6  & 21 & 31 & 18 & -- \\ \cline{2-6}
        & P9  & 41 & 42 & 21 & -- \\ \cline{2-6}
        & P11 & 64 & -- & -- & -- \\ \cline{2-6}
        & P13 & 42 & 40 & 28 & -- \\ \cline{2-6}
        & P16 & 57 & 54 & 30 & -- \\ \hline
        \hline

        \multirow{10}{*}{\textbf{PQS}} 
        & P3  & 82 & 23 & -- & -- \\ \cline{2-6}
        & P4  & 98 & -- & -- & -- \\ \cline{2-6}
        & P7  & 48 & 29 & 23 & -- \\ \cline{2-6}
        & P8  & 65 & 32 & -- & -- \\ \cline{2-6}
        & P10 & 46 & 10 & 29 & -- \\ \cline{2-6}
        & P12 & 73 & 31 & 15 & -- \\ \cline{2-6}
        & P14 & 55 & 30 & 26 & -- \\ \cline{2-6}
        & P15 & 56 & 23 & -- & -- \\ \hline

    \end{tabular}
\end{table}

\end{document}